\documentclass{iau}

\usepackage{amsmath}
\usepackage{graphicx}
\usepackage{multirow}
\usepackage{xspace} 
\usepackage{color, soul} 

\usepackage{xcolor}
\usepackage[normalem]{ulem}

\definecolor{seagreen}{rgb}{0.190, 0.525, 0.361}

\newcommand{\eq}[2][]{\begin{align}
                          #2
\end{align}}

\newcommand{\BH}{\mathrm{BH}}

\newcommand{\units}[1]{\,\mathrm{#1}}

\newcommand{\Gaia}{{\it Gaia}\xspace}
\newcommand{\gb}{\Gaia~BH3\xspace}
\newcommand{\feh}{\mathrm{[Fe/H]}}
\newcommand{\sbh}{S-BH\xspace}

\def\subinrm#1{\sb{\rm#1}}
    {\catcode`\_=13 \global\let_=\subinrm}
\mathcode`\_="8000
\def\upsubscripts{\catcode`\_=12 }

\upsubscripts

\newcommand{\msun}{{\rm M}_\odot}

\renewcommand{\eqref}[1]{(equation~\ref{#1})}

\begin{document}

\lefttitle{Daniel Marín Pina}
\righttitle{The formation of \textit{Gaia} BH3}

\jnlPage{1}{7}
\jnlDoiYr{2021}
\doival{10.1017/xxxxx}

\aopheadtitle{Proceedings IAU Symposium}
\editors{C. Sterken,  J. Hearnshaw \&  D. Valls-Gabaud, eds.}

\title{The formation of \textit{Gaia} BH3}

\author{Daniel Marín Pina$^{1,2}$, Mark Gieles$^{1,3}$, Sara Rastello$^{1,2}$, and Giuliano Iorio$^{1,2}$}

\affiliation{$^{1}$ Institut de Ciències del Cosmos (ICCUB), Martí i Franquès 1, 08028 Barcelona, Spain}
\affiliation{$^{2}$Departament de F\'isica Qu\`antica i Astrof\'isica (FQA), Universitat de Barcelona (UB), Mart\'i i Franqu\`es 1, 08028 Barcelona, Spain}
\affiliation{$^{3}$ICREA, Pg. Llu\'is Companys 23, 08010 Barcelona, Spain}

\begin{abstract}
The \textit{Gaia} collaboration announced the discovery of a massive black hole (BH) with a low-mass giant star companion, \gb, located in the ED-2 stellar stream. The properties of \gb bridge the gap between known Milky Way BHs and extragalactic BHs found with gravitational waves (GWs). We aim to determine the most likely formation scenario for \gb in the progenitor cluster of the ED-2 stream. We perform $N$-body simulations of that progenitor cluster and find that, most likely, \gb formed from a stellar binary that formed during cluster formation, which then underwent multiple dynamical interactions that significantly altered its properties, including exchanging the companion star. We highlight the importance of cluster dynamics and discard a formation scenario where it evolved in quasi-isolation.
\end{abstract}


\maketitle


\section{Introduction}

Recently, we have begun to uncover a population of BHs that do not merge or accrete; they are `dormant' BHs. One of the methods to detect dormant BHs is in star-BH (S-BH) binaries via the astrometric movement of the star. In this regard, the \textit{Gaia} mission has been key to uncovering these systems, with a confirmed detection of three dormant BHs, with many more such systems expected to be discovered in the next \textit{Gaia} Data Release (DR4).

From the discovered dormant BHs, \gb stands out. It consists of a massive BH ($m_\BH = 33\units{\msun}$) in a wide, eccentric orbit with an extremely metal-poor, low-mass giant star. \gb is remarkably close to the Sun ($590\units{pc}$), in an orbit around the Milky Way (MW) halo associated with the ED-2 stellar stream \citep{Balbinot2023, Balbinot2024}, which formed from a now-dissolved star cluster. The BH is more massive than any other (stellar-origin) BH detected in the MW, and is close to the median BH mass detected via GWs \citep[see Fig.~3 in][]{MarinPina2024}. This binary bridges the populations of BHs detected via electromagnetic measurements and GWs, and understanding its formation pathway can provide constraints on the populations of binaries with BHs and the formation of GW sources.

There are two competing explanations for the origin of \gb: either it was formed from the quasi-isolated evolution of a binary of stars \citep[e.g.,][]{Iorio2024, ElBadry2024}, or it formed via dynamical interactions in the progenitor cluster of the ED-2 stream \citep{MarinPina2024}. In this contribution, we present models of the progenitor cluster of the ED-2 stream and show that the effect of dynamics in the formation of \gb can not be neglected, therefore favouring a dynamical over a quasi-isolated formation scenario. 

\section{Methods}
\label{sec:methods}
We show preliminary results of the simulations of the ED-2 progenitor cluster from Marín Pina et al. (in prep, MP25 hereafter). 
These simulations are run using the high-performance $N$-body code \textsc{petar} \citep{Wang2020}. We initialise all models using a \cite{Plummer1911} density profile with a \cite{Kroupa2001} initial mass function (IMF) between $0.08\units{\msun}$ and $150\units{\msun}$. 
We fix the (approximate) dissolution time of the cluster to $t_{dis}\in[4, 8]\units{Gyr}$, and then use different values of initial cluster mass and half-mass density within the constraints imposed by \cite{Balbinot2024}, $2\times 10^3 \lesssim M_0/\msun \lesssim 5.2\times 10^4$ (see MP25 and \citealt{GielesGnedin2023}).

These models are run both with and without a primordial binary population (originated as part of the cluster formation process). For the models with primordial binaries, we pair up all stars above $5\units{\msun}$ using the \cite{SanaEvans2011} period and eccentricity distributions. Then, we pair lower-mass stars using the \cite{DuquennoyMayor1991} period distribution and a thermal eccentricity distribution until we reach an initial binary fraction of $f_b=0.4$. 
We assume a metallicity equal to that of the \gb star, $\feh = -2.56$. 

We evolve the cluster until $13\units{Gyr}$, where single and binary stars are evolved with \textsc{bseEmp}  \citep{Tanikawa2020}.
 The cluster is evolved in a Milky Way-like potential using the \textsc{mwpotential2014} in \textsc{galpy} \citep{galpy}. The initial conditions are such that, after $13\units{Gyr}$, the centre of mass of the cluster matches the current position of \gb.

\section{Formation of the ED-2 stream}
\label{sec:stream}
To verify that our simulations are representative of the parent cluster of the ED-2 stream, we compare the streams formed in our simulations to the ED-2 \Gaia data reported in \cite{Balbinot2023}. Since the cluster is old \citep[$t\gtrsim 13\units{Gyr}$,][]{Balbinot2024}, it must have undergone several orbits around the Milky Way. Since there are many uncertainties associated with the evolution of the galactic potential, our goal is not to match the shape of the observed stream exactly, but rather to show that our models dissolve and form a stream similar to ED-2.

Because the cluster is dissolved, most stars are dispersed through the Galaxy and are faint. To account for detectability, we convert the output of our simulations to the \Gaia $G$ magnitudes using the corrections from \cite{Flower1996, Torres2010, Jordi2010}. We then limit our sample to stars above the $G$-band limits of the ED-2 sample from \cite{Balbinot2023}.

\begin{figure*}[htbp]
    \centering
    \includegraphics[width=\linewidth]{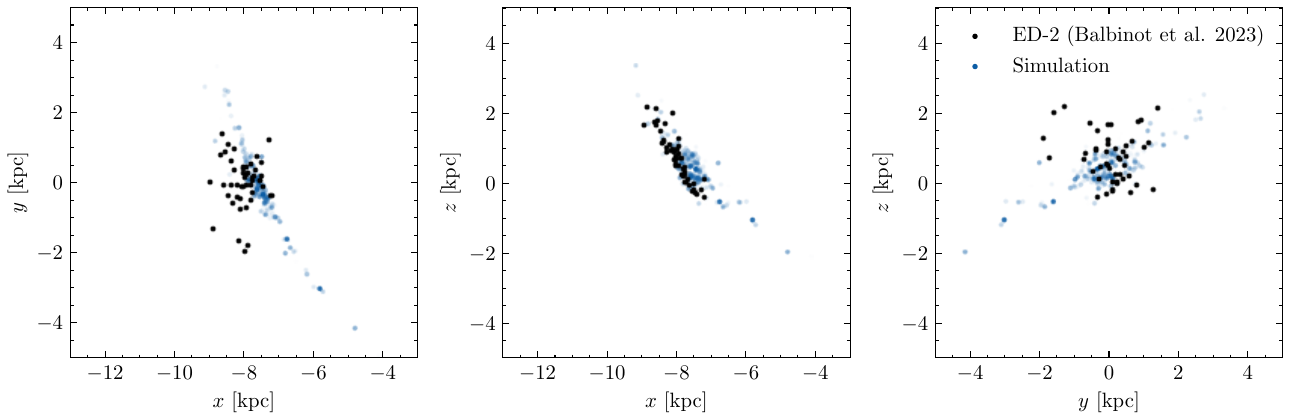}\vspace{-0.2cm}
    \caption{Positions of the stars of the ED-2 stream in Galactocentric cartesian coordinates. In black, \Gaia data from \cite{Balbinot2023}. In blue, a simulation with $M_0=7.7\times 10^3\units{\msun}$ and $\rho_{h, 0}=760 \units{\msun}\units{pc}^{-3}$, with each star coloured according to its magnitude.}
    \label{fig:stream}
\end{figure*}

In Fig.~\ref{fig:stream}, we show the result of one of our simulations that dissolved $5\units{Gyr}$ ago, and the members of ED-2 from \cite{Balbinot2023}. The data qualitatively match the output of the simulations, which shows that the simulations are representative of the ED-2 progenitor.

\vspace{-0.2cm}

\section{The origin of \gb}
\label{sec:gbh3}
In the previous section, we showed that our models of GCs dissolve in the \gb orbit and form streams similar to ED-2, therefore showing that \gb formed in the ED-2 progenitor cluster. In a cluster, we distinguish the following formation channels for \gb: 
\begin{itemize}
    \item `primordial', where the binary originated as part of the cluster formation process, but interactions with other stars, BHs, and binaries significantly altered its properties;
    \item `exchange', where the BH formed in a primordial binary but exchanged its companion star via dynamical interactions;
    \item `dynamical', where the binary was assembled from a previously-unbound star and a BH via dynamical interactions.
\end{itemize}

In the following Sections, we determine the most likely formation channel for \gb and explore whether dynamical interactions have a significant effect on the formation of \sbh systems in the ED-2 stream progenitor, thereby testing a quasi-isolated formation scenario for \gb. We stress, however, that all \sbh formed from primordial binaries in the cluster simulations undergo multiple strong interactions.


\subsection{Formation of \gb in the ED-2 progenitor}
\label{ssec:formation}
Here we estimate the likelihood of the different formation scenarios.
For the first two scenarios, we use the cluster models with primordial binaries, separating the \sbh depending on whether they were paired up at the beginning of the simulation. For the third formation scenario, we use our models without primordial binaries. The distribution of the parameters of \sbh in our simulations is shown in Fig.~\ref{fig:properties}.

To separate the most likely formation scenario, we compute the marginal probabilities (posteriors) $P(\text{prim}\,|\,\theta_G)$, $P(\text{exc}\,|\,\theta_G)$, and $P(\text{dyn}\,|\,\theta_G)$, with $\theta_G=\{\log P, e, q\}$ the measured parameters of \gb. Each posterior represents the probability that \gb formed in its respective scenario. Using Bayesian statistics, we have
\eq{P(\text{dyn}\,|\,\theta_G)=\frac{P(\theta_G\,|\,\text{dyn})P(\text{dyn})}{P(\theta_G\,|\,\text{prim})P(\text{prim})+P(\theta_G\,|\,\text{exc})P(\text{exc})+P(\theta_G\,|\,\text{dyn})P(\text{dyn})}}

Here, $P(\text{prim})$, $P(\text{exc})$, and $P(\text{dyn})$ are the respective priors. $P(\theta_G\,|\,\text{dyn})$ is the probability density function (PDF) of the parameters found in the dynamical channel, evaluated at $\theta_G$. The priors are computed from the total number of \sbh systems per cluster; the PDF is computed via a Kernel Density Estimation (KDE). The posteriors for the other channels are computed using the same approach.

On average, there are significantly more \sbh systems in clusters with primordial binaries. In general, $P(\text{prim})\gg P(\text{dyn})$ and $P(\text{exc})\gg P(\text{dyn})$. In our Bayesian analysis, the posteriors are $P(\text{prim}\,|\,\theta_G)=8\%$, $P(\text{dyn}\,|\,\theta_G)=3\%$, and $P(\text{exc}\,|\,\theta_G)=89\%$. We conclude that the most likely scenario is that \gb formed as an exchange \sbh binary.



\subsection{Could \gb have formed in quasi-isolation?}
\label{ssec:quasiiso}
In the previous Section, we showed that \gb most likely formed from a primordial binary in the ED-2 progenitor cluster that exchanged its companion star. We now test the effect that dynamical interactions in the cluster had on shaping the properties of the binaries. This is akin to asking if \gb could have formed in quasi-isolation, as proposed in e.g.~\cite{Iorio2024, ElBadry2024}.

\begin{figure*}[htbp]
    \centering
    \includegraphics[width=\linewidth]{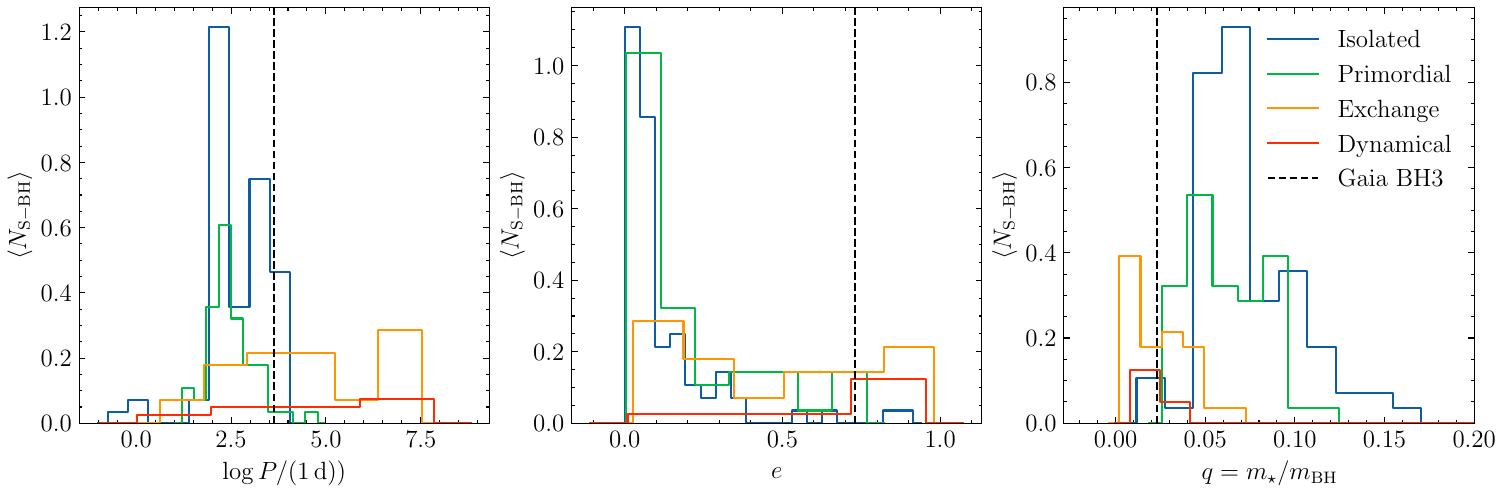}
    \caption{Average number of \sbh binaries per cluster in our simulations as a function of the logarithm of the period (left), the eccentricity (middle), and the mass ratio (right). In green and yellow, the primordial and exchanged (respectively) \sbh formed in simulations with primordial binaries; in red, the dynamical \sbh formed in simulations without primordial binaries; in blue, the \sbh formed from primordial binaries extracted from the simulations and evolved in isolation; in dashed black, the parameters of \gb.}
    \label{fig:properties}
\end{figure*}

To test the quasi-isolated channel, we have extracted all binaries at the start of our cluster simulations and evolved them in isolation. We have used the same binary stellar evolution code as in the cluster simulations, so that all differences are strictly due to dynamical interactions. In Fig.~\ref{fig:properties}, we show the properties of binaries evolved in isolation, as well as those found at the end of our cluster simulations.

As expected, we find that dynamics significantly modifies the \sbh binaries formed in the models. About $\sim 40\%$ of \sbh binaries in clusters have undergone at least one exchange of the companion star. From the binaries that do not exchange their companions, their distributions are skewed towards smaller periods and eccentricities. Furthermore, the \sbh binaries in our simulations undergo a median of 84 strong binary-single interactions and 9 strong binary-binary interactions. These data reject a quasi-isolated approach to studying the formation of \gb.

\section{Conclusions}
\label{sec:conclusions}
In this paper, we present part of the simulations of MP25.
We show that the likely progenitor of the ED-2 stream is a now-dissolved GC, with a mass within the uncertainties of \cite{Balbinot2024}. We show that \gb likely formed from a primordial binary of the ED-2 cluster that exchanged its companion star via dynamical encounters. We neglect the possibility that \gb formed as a dynamical binary from a previously-unbound BH and a star due to the low rates of this channel. Furthermore, we show that the properties of primordial \sbh binaries in the ED-2 progenitor cluster are modified by dynamical interactions. We, therefore, recommend being cautious when interpreting the comparison of isolated binary models to the observed properties of \gb.


\begin{acknowledgements}
    DMP and MG acknowledge financial support from the grants PRE2020-091801, PID2021-125485NB-C22, PID2024-155720NB-I00, CEX2019-000918-M, CEX2024-001451-M funded by MCIN/AEI/10.13039/501100011033 (State Agency for Research of the Spanish Ministry of Science and Innovation). 
\end{acknowledgements}

\bibliographystyle{iaulike} 
\bibliography{bibliography} 

\end{document}